\documentclass[12pt]{article}
\usepackage{amsmath,amssymb,amsthm,mathrsfs,url,hyperref}
\usepackage{color}

\title{Limitations to Genuine Measurements in Ontological Models of Quantum Mechanics}

\author{
Roderich Tumulka\footnote{Fachbereich Mathematik, Eberhard-Karls-Universit\"at, 
	Auf der Morgenstelle 10, 72070 T\"ubingen, Germany. 
	E-mail: roderich.tumulka@uni-tuebingen.de}
}

\date{September 8, 2022}

\newcommand{\Hilbert}{\mathscr{H}}
\newcommand{\be}{\begin{equation}}
\newcommand{\ee}{\end{equation}}
\newcommand{\scp}[2]{\langle #1|#2\rangle}

\newcommand{\RRR}{\mathbb{R}}
\newcommand{\SSS}{\mathbb{S}}
\newcommand{\CCC}{\mathbb{C}}
\newcommand{\PPP}{\mathbb{P}}
\newcommand{\Values}{\Omega}
\newcommand{\Valg}{\mathscr{F}} 
\newcommand{\cEXP}{\mathcal{EXP}}
\newcommand{\cPOVM}{\mathcal{POVM}}
\newcommand{\sE}{\mathscr{E}}
\newcommand{\sL}{\mathscr{L}}
\newcommand{\sG}{\mathscr{G}}

\theoremstyle{plain}
\newtheorem{thm}{Theorem}
\theoremstyle{definition}
\newtheorem{defn}{Definition}
\newtheorem{ex}{Example}

\begin{document}
\maketitle
\begin{abstract}
Given an ontological model of a quantum system, a ``genuine measurement,'' as opposed to a quantum measurement, means an experiment that determines the value of a beable, i.e., of a variable that, according to the model, has an actual value in nature before the experiment. We prove a theorem showing that in every ontological model, it is impossible to measure all beables. Put differently, there is no experiment that would reliably determine the ontic state. This result shows that the positivistic idea that a physical theory should only involve observable quantities is too optimistic.

\medskip

\end{abstract}

\section{Introduction}

``Beables'' is a word coined by John Bell \cite{Bell76b} for variables that have values in nature even when nobody looks. The ``ontic state'' of a quantum system \cite{Spe04b,Lei14} means the actual, factual, physical state in reality. The ontic state could be described through the values of all beables. An ``ontological model'' of quantum mechanics \cite{Spe04b,Lei14} is a proposal for what the ontic states might be and which laws might govern them. We denote an ontological model by $M$ (definition below), the ontic state by $\lambda$, and the set of ontic states by $\Lambda$.

We prove a theorem asserting that, given any ontological model $M$ that is empirically adequate (i.e., whose observable predictions agree with the rules of quantum mechanics) and covers sufficiently many ``observables,'' there is no experiment that would determine $\lambda$. Here, we take the ontological model $M$ as known and only ask whether an experiment would reveal $\lambda$ in a (hypothetical) world governed by $M$.

Our result goes against some kinds of positivism, as it poses a \emph{limitation to knowledge}: It shows that not everything that exists in reality is observable, and not every variable that has a value can be measured. 
Various theories of quantum mechanics have been criticized for postulating the real existence of objects (or variables) that are not fully observable (respectively, measurable); in view of our result, this criticism appears inappropriate because every theory of quantum mechanics will have this property.

Our proof makes use of the concept of POVM (positive-operator-valued measure\footnote{A \emph{POVM on the set $\Values$ acting on the Hilbert space $\Hilbert$} is defined to be an association $E$ of a positive operator $E(B)$ on $\Hilbert$ with every (measurable) subset $B$ of $\Values$ such that $E$ is countably additive on disjoint subsets and $E(\Values)=I$. For more detail, see, e.g., \cite{povm} or \cite[Sec.\ 5.1]{book}.}) and of the \emph{main theorem about POVMs} (see \cite{DGZ04} or \cite[Sec.\ 5.1]{book}), which asserts that {\it for every experiment $\sE$ on a quantum system $S$ whose possible outcomes lie in a set $\Values$ (equipped with sigma algebra\footnote{A \emph{sigma algebra on a set $\Values$} is defined to be a collection $\Valg$ of subsets of $\Values$ that is closed under the operations of finite or countable union, finite or countable intersection, and complement. Its elements are called the ``measurable subsets'' of $\Values$. Measures (such as volume, probability, or POVMs) are defined on a sigma algebra.} $\Valg$), there exists a POVM $E$ on $(\Values,\Valg)$ such that, whenever $S$ has wave function $\psi$ at the beginning of $\sE$, the random outcome $Z$ has probability distribution given by}
\begin{equation}\label{PPPPOVM}
\PPP(Z \in B) = \scp{\psi}{E(B)|\psi}~~~~\forall B \in \Valg\,.
\end{equation}

We emphasize that in contrast to the concept of an ideal quantum measurement of a self-adjoint operator, which corresponds only to very special experiments, the concept of a POVM covers, according to the main theorem about POVMs, \emph{all possible} experiments that could in principle be performed on a single given quantum system $S$.

\section{Definitions}

Let $\SSS(\Hilbert)=\{\psi\in\Hilbert: \|\psi\|=1\}$ denote the unit sphere of the Hilbert space $\Hilbert$ and $\cPOVM$ the set of all POVMs acting on $\Hilbert$. 

\begin{defn}
An \emph{ontological model} $M$ of a quantum system with Hilbert space $\Hilbert$ consists of 
\begin{itemize}
\item[(i)] a set $\Lambda$ called the ontic space with sigma algebra $\sL$; 
\item[(ii)] for every $\psi\in\SSS(\Hilbert)$, a probability measure $\varrho^\psi$ over $(\Lambda,\sL)$;\footnote{One could imagine that different procedures for preparing $\psi$ lead to different distributions over $\Lambda$; for our purposes, we can simply choose one such distribution for each $\psi$.}
\item[(iii)] an index set $\cEXP$ representing the set of possible experiments; 
\item[(iv)] a mapping $E:\cEXP\to\cPOVM$ associating with every experiment the POVM according to the main theorem about POVMs;\footnote{In essence, it is assumed here that the experiments in $\cEXP$ are \emph{physically possible}.}
\item[(v)] for every ontic state $\lambda\in\Lambda$ and every experiment $\sE\in\cEXP$, a probability distribution $P_{\lambda,\sE}$ over the value space $\Values$ of $\sE$ (thought of as the distribution of the outcome when $\sE$ is applied to a system in state $\lambda$).\footnote{For mathematicians: $P_{\lambda,\sE}(B)$ is required to be a measurable function of $\lambda$ for every $\sE$ and $B$.}
\end{itemize}
\end{defn}

\begin{ex}
Bohmian mechanics \cite{Bohm52a,DGZ13,book} is an ontological model for $\Hilbert=L^2(\RRR^{3N})$ with ontic state given by the pair $\lambda=(Q,\Psi)$ consisting of the configuration $Q$ and the wave function $\Psi$, so (i)~$\Lambda=\RRR^{3N}\times \SSS(\Hilbert)$ for $N$ particles. (ii)~The probability measure over $\Lambda$ is
\be\label{BMrho}
\varrho^{\psi}(dQ \times d\Psi) = |\psi(Q)|^2 \,\delta(\Psi-\psi) \, dQ \, d\Psi\,.
\ee
(iii)~An experiment $\sE$ can be specified by providing the Hilbert space $\Hilbert_A=L^2(\RRR^{3M})$ of the apparatus, the wave function $\phi_A \in \SSS(\Hilbert_A)$ of the apparatus at the beginning of $\sE$, the Hamiltonian $H$ of $S$ and $A$ together during $\sE$, the duration $T$ of $\sE$, and the ``calibration function'' $\zeta:\RRR^{3M}\to \Values$ that extracts the outcome $Z$ from the configuration of $A$. (iv)~Then \cite{DGZ04,book}
\be\label{BME}
E_\sE(B) := \scp{\phi_A}{e^{iHT} [I\otimes P_A(\zeta^{-1}(B))] e^{-iHT}|\phi_A}
\ee
defines a POVM $E_\sE$ on $\Values$ acting on $\Hilbert$, where the inner product is the partial inner product in $\Hilbert_A$, $I$ is the identity operator on $\Hilbert$, and $P_A$ is the position POVM of $A$, i.e., $P_A(B_A)$ is the multiplication operator by the characteristic function of the set $B_A\subseteq \RRR^{3M}$. (v)~To define $P_{\lambda,\sE}$ for $\lambda=(Q,\Psi)$, solve the Schr\"odinger equation with $H$ and initial datum $\Psi\otimes\phi_A$ and Bohm's equation of motion with initial datum $(Q,Q_A)$, where $Q_A$ is random with distribution $|\phi_A|^2$. Call the solution $(Q_S(t),Q_A(t))$. Then
\be\label{BMP}
P_{\lambda,\sE}(B):=\PPP\bigl(\zeta(Q_A(T))\in B\bigr)~~~~\forall B \in\Valg\,.
\ee
It is known (e.g., \cite[Sec.~7.2]{DGZ04} and \cite[Sec.~5.1.3.]{book}) that inhabitants of a Bohmian world can measure $Q$ but not $\Psi$ (so it is ironic that $Q$ gets called a ``hidden variable'').
\end{ex}

\begin{defn}
An ontological model $M$ is said to be \emph{empirically adequate} if and only if for every $\psi\in\SSS(\Hilbert)$, 
every $\sE\in\cEXP$, and every $B\in\Valg$,
\be\label{adequate}
\int_\Lambda \varrho^\psi(d\lambda) \, P_{\lambda,\sE}(B) = \scp{\psi}{E_{\sE}(B)|\psi} \,.
\ee
\end{defn}

\begin{ex}
Bohmian mechanics is empirically adequate; in fact, it is well known that its observable predictions agree with the rules of quantum mechanics. Specifically, the calculation verifying \eqref{adequate} with \eqref{BMrho}, \eqref{BME}, and \eqref{BMP} can be found in \cite[Sec.~5]{DGZ04} and \cite[Sec.~5.1.2]{book}.
\end{ex}

For any 1d subspace $g$ of $\Hilbert$, let $P_g$ denote the projection to $g$ (i.e., $P_g=|\psi\rangle\langle\psi|$ for $g=\CCC\psi$ with $\|\psi\|=1$) and $F_g$ the POVM on $\{0,1\}$ given by
\be\label{Fg}
F_g(\{1\})=P_g~~~\text{and}~~~F_g(\{0\})=I-P_g
\ee
with $I$ the identity operator.

\begin{defn}
We call an ontological model $M$ \emph{line complete} if for every 1d subspace $g$ of $\Hilbert$, there is an experiment $\sE\in\cEXP$ with $E_{\sE}=F_g$.
\end{defn}

\begin{ex}
For spin space $\Hilbert=\CCC^2$ and $\psi\in\SSS(\Hilbert)$, the Stern--Gerlach experiment with magnet oriented in the direction $\psi^\dagger \boldsymbol{\sigma} \psi$ in physical 3-space and outcome ``up'' (``down'') represented by $Z=1$ ($Z=0$) has POVM $F_{\CCC\psi}$. Here, $\boldsymbol{\sigma}=(\sigma_x,\sigma_y,\sigma_z)$ is the triple of Pauli matrices.
\end{ex}

\section{Result}

Our claim is that it is impossible to measure $\lambda$. That is, there is no experiment $\sG\in\cEXP$ that, when applied to a system in state $\lambda$, yields $\lambda$ as the outcome. Formally:

\begin{thm}
Given a Hilbert space $\Hilbert$ with $\dim\Hilbert\geq 2$ and an empirically adequate, line complete ontological model $M$ for $\Hilbert$, there is no $\sG\in\cEXP$ such that $(\Values,\Valg)=(\Lambda,\sL)$ and
\be\label{measurelambda}
P_{\lambda,\sG}(B) = 1_{\lambda\in B}~~~~\forall B\in\Valg\,.
\ee
\end{thm}

Equivalently, \eqref{measurelambda} can be rewritten as $P_{\lambda,\sG} = \delta_{\lambda}$ with the notation $\delta_{\lambda}$ for the normalized measure concentrated in the single point $\lambda$. A key step toward Theorem~1 is the following statement:

\begin{thm}
Given a Hilbert space $\Hilbert$ with $\dim\Hilbert\geq 2$ and an empirically adequate, line complete ontological model $M$ for $\Hilbert$,
there is no POVM $G$ on $(\Lambda,\sL)$ acting on $\Hilbert$ such that
\be\label{psiG}
\varrho^\psi(A) = \scp{\psi}{G(A)|\psi}
\ee
for all $A\in\sL$ and $\psi\in\SSS(\Hilbert)$.
\end{thm}

\section{Proofs}

\begin{proof}[Proof of Theorem 1 from Theorem 2]
If such a $\sG\in\cEXP$ existed, it would be associated with a POVM $G=E_\sG$ on $(\Lambda,\sL)$ acting on $\Hilbert$. By \eqref{adequate} for $\sE=\sG$,
\be
\int_\Lambda \varrho^\psi(d\lambda) \, P_{\lambda,\sG}(B) = \scp{\psi}{G(B)|\psi}
\ee
for all $\psi\in\SSS(\Hilbert)$ and $B\in\Valg$. By \eqref{measurelambda}, the left-hand side equals
\be
\int_\Lambda \varrho^\psi(d\lambda) \, 1_{\lambda\in B} = 
\int_B \varrho^\psi(d\lambda) =  \varrho^\psi(B)\,,
\ee
so
\be
\varrho^\psi(B)= \scp{\psi}{G(B)|\psi}
\ee
for all $\psi\in\SSS(\Hilbert)$ and $B\in\Valg=\sL$, which is impossible by Theorem 2.
\end{proof}

Before we give the full proof of Theorem 2, let us give an outline. We will first deduce that for every experiment $\sE$, $E_\sE$ is the $G$-average of $P_{\lambda,\sE}$. Then we consider yes/no experiments corresponding to 1d subspaces $g$ of $\Hilbert$ and the regions $\Lambda_g$ in the ontic space where the probability of a \emph{yes} answer is non-zero. We find that $\Lambda_g$ and $\Lambda_h$ must be disjoint (up to $G$-null sets) whenever $g$ and $h$ are distinct, even if they are not orthogonal. This leads to a conflict with the requirement that, for any subset $S\in\sL$ of $\Lambda$, $G(S)\leq G(\Lambda)=I$.

\begin{proof}[Proof of Theorem 2]
We assume that such a $G$ exists and will derive a contradiction. Putting \eqref{adequate} and \eqref{psiG} together, we obtain that
\be
\int_\Lambda \scp{\psi}{G(d\lambda)|\psi} \, P_{\lambda,\sE}(B) = \scp{\psi}{E_{\sE}(B)|\psi}
\ee
for all $B\in\Valg$. The left-hand side can be re-written as
\be
\Bigl\langle \psi \Big| \int_\Lambda G(d\lambda) \, P_{\lambda,\sE}(B) \Big| \psi \Bigr\rangle\,.
\ee
Since $\scp{\psi}{R|\psi}=\scp{\psi}{S|\psi}$ for all $\psi\in\SSS(\Hilbert)$ only if $R=S$ (by the polarization identity), we have that
\be
\int_\Lambda G(d\lambda) \, P_{\lambda,\sE}(B) = E_{\sE}(B)
\ee
for all $B\in\Valg$. 
Since $M$ is assumed to be line complete, there is, for every 1d subspace $g$ of $\Hilbert$,  
an experiment $\sE(g)\in \cEXP$ so that $E_{\sE(g)}=F_g$ as in \eqref{Fg}. For $B=\{1\}$, we obtain that
\be\label{intg}
\int_\Lambda G(d\lambda) \, P_{\lambda,\sE(g)}(\{1\}) = P_g \,.
\ee
Note that for every $g$, $P_{\lambda,\sE(g)}(\{1\})$ is a function of $\lambda$ with values in $[0,1]$, and define
\be
\Lambda_g:=\{\lambda\in\Lambda:P_{\lambda,\sE(g)}(\{1\}) >0\}\,.
\ee
It follows that $P_{\lambda,\sE(g)}(\{1\})\leq 1_{\Lambda_g}(\lambda)$ and thus
\be\label{GPg}
G(\Lambda_g) = \int_\Lambda G(d\lambda) \, 1_{\Lambda_g}(\lambda) \geq
\int_\Lambda G(d\lambda) \, P_{\lambda,\sE(g)}(\{1\}) = P_g \,.
\ee

On the other hand, for any set $A\in\sL$ with $A\subseteq \Lambda_g$, $G(A)$ must be a non-negative multiple of $P_g$: indeed, $G(A)$ is a positive operator, and if $G(A)$ had any eigenvector $\notin g$ with nonzero eigenvalue, then there would exist $0\neq \chi\in g^\perp$ with $\scp{\chi}{G(A)|\chi}>0$ and so 
\be
\Bigl\langle \chi \Big| \int_\Lambda G(d\lambda) \, P_{\lambda,\sE(g)}(\{1\}) \Big| \chi \Bigr\rangle >0,
\ee
in contradiction to \eqref{intg} together with $\scp{\chi}{P_g|\chi}=0$.

Since $G(A)$ is a multiple of $P_g$,
\be
0\leq G(A)\leq P_g \,.
\ee
Two consequences: First, by \eqref{GPg},
\be
G(\Lambda_g)=P_g\,.
\ee
Second, for 1d subspaces $g\neq h$, $\Lambda_g$ and $\Lambda_h$ must be disjoint up to $G$-null sets,
\be
G(\Lambda_g \cap \Lambda_h)=0
\ee
(because this operator must be $\leq P_g$ and $\leq P_h$, and the only positive operator achieving that is 0).

Now let $\psi_1$ and $\psi_2$ be mutually orthogonal unit vectors, set $\psi_3=\frac{1}{\sqrt{2}}(\psi_1+\psi_2)$ and $g_i=\CCC\psi_i$ for $i=1,2,3$. Then
\be
G\bigl(\Lambda_{g_1}\cup \Lambda_{g_2} \cup \Lambda_{g_3}\bigr)= 
G(\Lambda_{g_1})+ G(\Lambda_{g_2}) + G(\Lambda_{g_3})= 
P_{g_1}+P_{g_2}+P_{g_3}\,,
\ee
which has eigenvalue 2 in the direction of $\psi_3$, in contradiction to $G(S)\leq G(\Lambda)=I$ for every $S\in\sL$.
\end{proof}

\section{Concluding Remarks}

The result is robust against small perturbations in the sense that even an \emph{approximate} measurement of $\lambda$ is impossible in an ontological model that is \emph{approximately} empirically adequate and contains experiments whose POVMs are \emph{approximately} $F_g$. After all, the final contradiction consisted in the fact that a certain eigenvalue that should be $\leq 1$ actually equals 2; if the eigenvalue were not exactly 2 but merely close to 2, it would still lead to a contradiction.

With regards to the broader significance, the result is in line with a sentiment expressed by
Bell \cite[Sec.~1]{Bell87a}:
\begin{quote}
To admit things not visible to the gross creatures that we are is, in my opinion, to show a decent humility, and not just a lamentable addiction to metaphysics.
\end{quote}
The question of limitations to knowledge has been addressed for various specific interpretations (e.g., \cite{Bohm52a,CT16,book}). It is also known that \emph{every} interpretation of quantum mechanics must entail limitations to knowledge. (See \cite[Sec.~5.1]{book}; the earliest proofs allowing this conclusion were given in \cite{Tum98} (translated and simplified in \cite{EE}) and \cite{Spe04b}.) One way of arriving at this conclusion is to note that wave functions cannot be measured \cite{DGZ04} and that the Pusey--Barrett--Rudolph theorem \cite{PBR,Lei14} shows that, under reasonable assumptions on the ontological model, the wave function is a beable.

Compared to these previous results, however, our present result addresses the question in a particularly direct way.

\bigskip

\noindent{\it Data Availability Statement}: Not applicable

\end{document}